\documentclass[aps,pra,reprint,superscriptaddress]{revtex4-1}

\usepackage{lineno}
\usepackage{hyperref}
\usepackage{graphicx}  % needed for figures
\usepackage{dcolumn}   % needed for some tables
\usepackage{bm}        % for math
\usepackage{amssymb}   % for math
\usepackage{epstopdf}
\usepackage{amsmath}
\usepackage{xspace}
\usepackage{verbatim}
\usepackage{color}
\usepackage{xfrac}
\usepackage{bbold}
\usepackage{soul}

\DeclareMathOperator{\Tr}{Tr}

\newcommand{\bra}[1]{\left\langle #1 \right|}
\newcommand{\ket}[1]{\left| #1 \right\rangle}

\hypersetup{
    colorlinks=true,linkcolor=blue,citecolor=blue,
    filecolor=blue,urlcolor=blue,breaklinks=true
}

\begin{document}
\title{Estimation of an Optomechanical Parameter via Weak Value Amplification}

\author{Sergio Carrasco}
\affiliation{Instituto de F\'{i}sica, Pontificia Universidad Cat\'{o}lica de Chile,
Casilla 306, Santiago, Chile}

\author{Miguel Orszag}
\email{morszag@fis.puc.cl}
\affiliation{Instituto de F\'{i}sica, Pontificia Universidad Cat\'{o}lica de Chile,
Casilla 306, Santiago, Chile}
\affiliation{Centro de \'{O}ptica e Informaci\'{o}n Cu\'{a}ntica, Camino la Pir\'{a}mide 5750, Huechuraba, Santiago, Chile}

\date{\today}

\begin{abstract}
In this article we present an experimental proposal for the estimation of an optomechanical parameter in the presence of noise. The estimation is based on the technique of weak value amplification which can enlarge the radiation pressure effect of a single photon on a mechanical oscillator.  In our setup we show that the weak value amplification technique is preferable for the estimation over a method that relies on a strong measurement with postselection, because the first method does not require a good prior knowledge of the parameter we wish to estimate,  while both strategies reach the same level of precision from a Fisher information perspective.  In the presence of strongly correlated noise the weak value amplification method is preferable, from a Fisher information perspective, than a standard measurement strategy that does not employ postselection and that is affected by the same type of noise. 
\end{abstract}

\maketitle

\section{Introduction}

Quantum metrology deals with the study of the lower bounds on measurement errors achievable through the use of quantum effects, and concerns with the features of the strategies employed to reach these bounds.  According to the central limit theorem, the average of $n$ independent measurements has an error that scales as $n^{-1/2}$. This behaviour, in the context of quantum mechanics,  has been called the standard quantum limit (SQL). Notably, with the use of quantum resources, e.g. quantum correlations of maximally path entangled NOON states \cite{Sanders1995} or squeezing \cite{Caves1981}, this limit can be surpassed, moving towards a Heisenberg type of scaling as $n^{-1}$ \cite{Giovannetti,Jarzyna}. 

In this article we investigate the precision of quantum parameter estimation in a cavity optomechanical system. The field of cavity optomechanics \cite{Aspelmeyer2014,Optomechanics} explores the interaction of light and mechanical motion through the radiation pressure. From a fundamental perspective, this type of systems allows the study of quantum effects in macroscopic objects, such as macroscopic superpositions and decoherence  \cite{Bose1999,Marshall2003,PepperGhobadi2012,PepperJeffrey2012} or optomechanical entanglement \cite{Marinkovic2018}. On the more practical side, optomechanical systems have been studied for force sensing applications\cite{Rugar2004,Jensen2008,Krause2012}, radiation pressure cooling \cite{Marquardt2007,WilsonRae2007,Genes2008}, gravitational wave detection \cite{Arva2013} and quantum information science applications \cite{Mancini2003}.

In our work we employ the technique of  weak value amplification (WVA) \cite{AAV1988,Svensson2013,TamirCohen2013,Kofman2012,Dressel2014} for the estimation of an optomechanical parameter. This method is based on \emph{weak measurements} of an observable together with a probabilistic procedure called \emph{postselection}, by which a final state of the system being measured is also specified (typically only the initial system state is specified). The ensemble average of this type of measurements corresponds to (the real part of) a weak value. Weak values may be larger than any of the eigenvalues of the observable being measured and therefore the effect on the measurement device can be amplified.  Weak values have been successfully applied to high precision estimation of small parameters, such as beam deflections \cite{Hall2008,Dixon2009,Starling2009}, frequency shifts \cite{StarlingFrequency2010}, phase shifts \cite{StarlingPhase2010},  Doppler shifts \cite{Viza2013}, longitudinal phase shifts \cite{Brunner2010}, angular rotations \cite{Loaiza2014} or temperature shifts \cite{Egan2012}. Other applications of weak measurements and weak values have been proposed for quantum control \cite{Coto2017} and for the generation of non classical states of macroscopic objects \cite{Montenegro2017,Carrasco2018}. 

The precision offered by measurements with postselection for parameter estimation has been analyzed in \cite{Feizpour2011,Kedem2012,Ferrie2013,Jordan2014,Sinclair2017} from the perspective of the signal to noise ratio (SNR) or the Fisher information.  In general, the precision does not improve with postselection as compared with a standard measurement without postselection. However, it is interesting to point out that the same level of precision can by reached with fewer information (because the postselection procedure reduces the size of the data set).  

In this article we propose an experimental setup for the estimation of an optomechanical parameter. 
The noise model taken into consideration is the one presented in \cite{Feizpour2011,Sinclair2017}, which allows to consider both a scenario affected by uncorrelated noise or strongly correlated noise. Our analysis of  the precision is based on the classical Fisher information. We compare the WVA method with a strong measurement with postselection, showing that both methods can reach the same level of precision but the first technique does not require a precise knowledge regarding the small parameter we wish to estimate. This conclusion holds when both measurements are affected either by uncorrelated or correlated noise.  

The WVA method is also compared with a standard measurement strategy which does not employ postselection. Both methods can reach the same level of precision in the presence of white noise, but the estimator constructed with the WVA technique relies on fewer observations of the measurement device. In the presence of strongly correlated noise, the WVA method offers a larger Fisher information than a standard measurement (without postselection) affected by correlated noise. Nevertheless, the (increased) Fisher information in the WVA method with correlated noise is smaller than the Fisher information in the uncorrelated noise case. 

This article is organized as follows. In Sec. \ref{sec:sec2} we describe the experimental setup proposed for the estimation of the optomechanical parameter. In Sec. \ref{sec:sec3} we present our analysis of the precision of the estimation, which is based on the classical Fisher information. Finally, in Sec. \ref{sec:sec4}, the results are summarized and commented.

\section{Experimental setup for parameter estimation}\label{sec:sec2} 

\begin{figure}
 \centering \includegraphics[width=\linewidth]{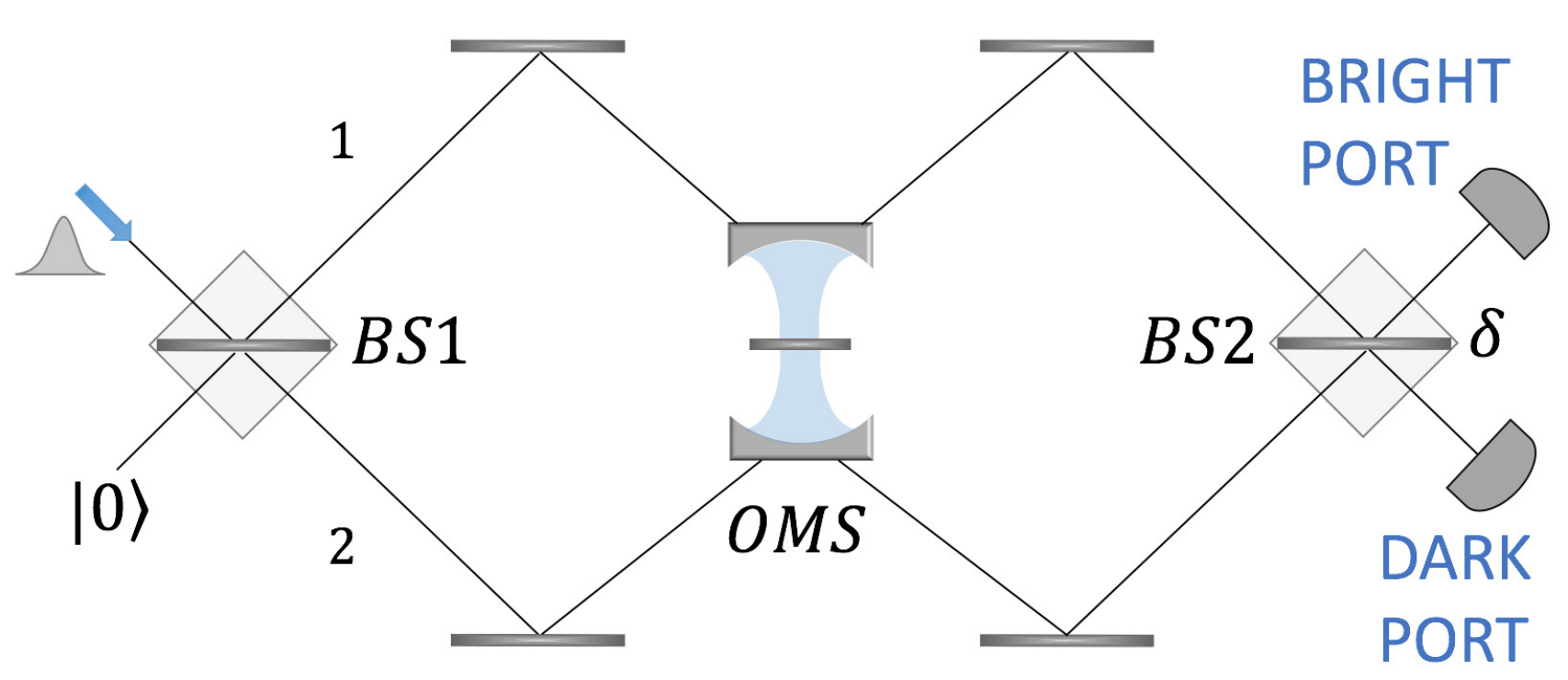}
 \caption{I1: Interferometer used to prepare the quantum state of the mechanical oscillator by performing postselection of photons. A mechanical oscillator is placed inside an optical cavity and prepared in the ground state. Photons are postselected in the dark port, which triggers the operation of I2, which is operated to observe of the average position of the mechanical oscillator.  
 } \label{fig:fig1}
\end{figure}

\begin{figure}
 \centering \includegraphics[width=\linewidth]{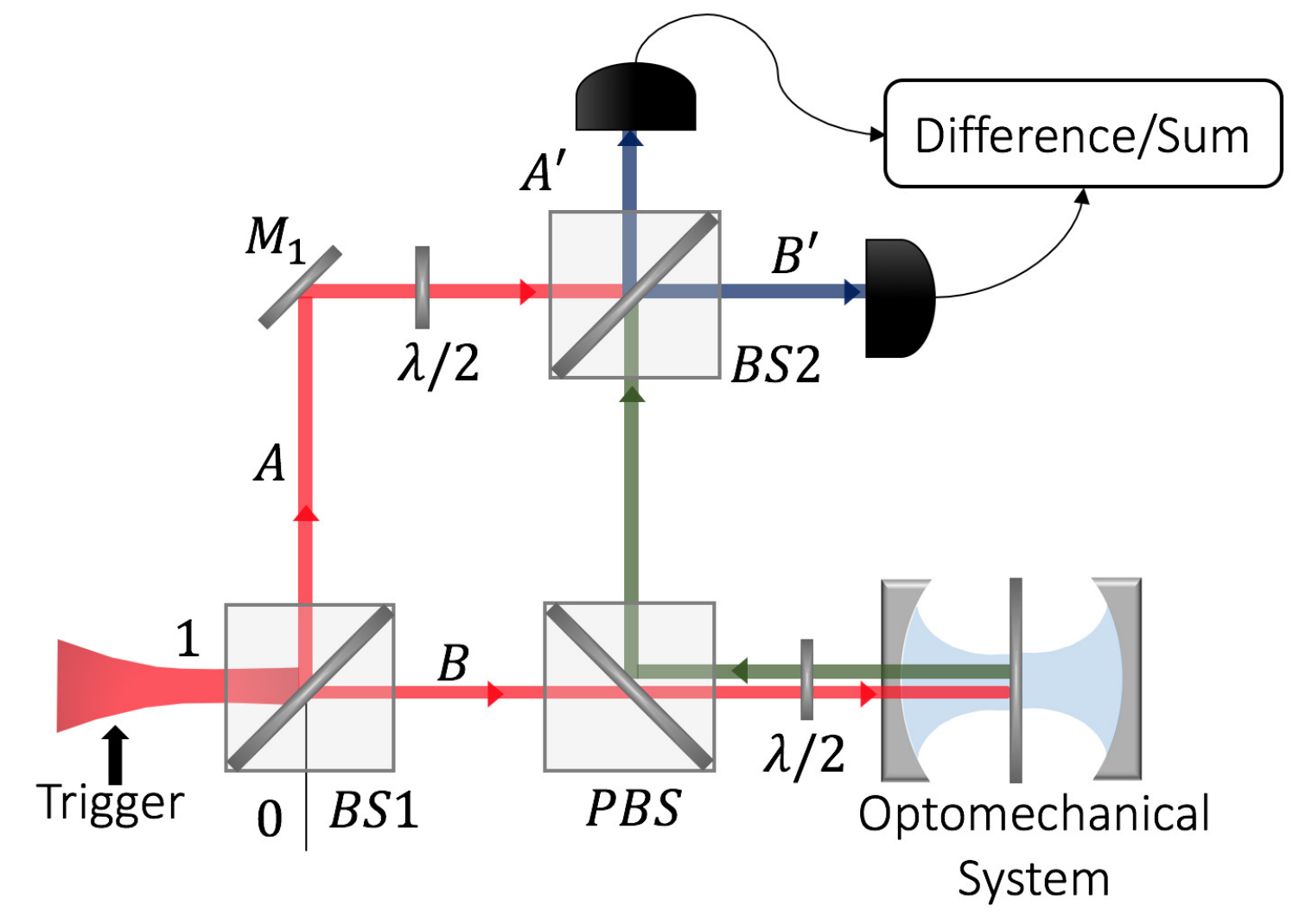}
 \caption{I2: Interferometric setup used to measure the length difference between both paths. A laser beam is triggered by a signal coming from I1. The laser is split into two paths, $A$ and $B$. The signal travelling through path $B$ acquires a phase which carries the information of the mechanical displacement (its average position). The sum and difference of photon counts at the detectors $A'$ and $B'$ allow to perform a measurement of the relative phase between both paths and thus of the mechanical displacement.} \label{fig:fig2}
\end{figure}

We will focus on the estimation of an optomechanical parameter that appears in the optomechanical interaction, the interaction between ``light and matter''. This task will be achieved using two interferometers, which will be referred as \emph{interferometer 1} (I1) and  \emph{interferometer 2} (I2). I1 and I2 are shown in Fig. \ref{fig:fig1} and \ref{fig:fig2}, respectively. 

I1 is a Mach-Zehnder interferometer with an optomechanical system along its arms. The task of I1 is to prepare the quantum state of a mechanical oscillator, while the aim of I2 is to perform an optical measurement of its average position. In this section we will describe the operation of I1, while I2 will be described in Sec. \ref{sec:sec3}. First, we will describe the optomechanical system in Sec. \ref{sec:sec2:sub1} and then the operation of I1, as a whole, in Sec. \ref{sec:sec2:sub2}.  

\subsection{Optomechanical sytem}\label{sec:sec2:sub1} 

\begin{figure}
 \centering \includegraphics[width=0.35\textwidth]{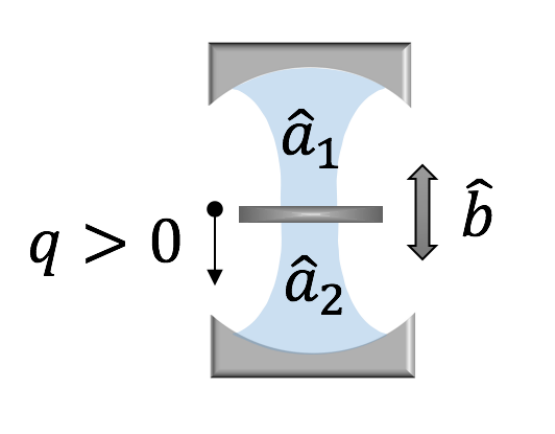}
 \caption{OMS: A vibrating mirror is placed in the middle of a Fabry-P\'{e}rot cavity. The fields interacts with the mirror via the radiation pressure force $\widehat{F}=\hbar (\omega/L)(\hat{a}^{\dagger}_1\hat{a}_1-\hat{a}^{\dagger}_2\hat{a}_2)$, where $\omega$ is the frequency (equal for both optical modes) and $L$ is the length of each side of the cavity. In our model there is no photon hopping between both sides.
 } \label{fig:fig3}
\end{figure}

An optomechanical system (OMS), illustrated in Fig.  \ref{fig:fig3}, is put inside a Mach-Zehnder interferometer, according to the configuration shown in Fig.  \ref{fig:fig1}. The OMS consists of an optical cavity with a high-Q vibrating mirror put in the middle (a micrometric mirror sustained by a cantilever). 

The hamiltonian that describes the OMS corresponds to
\begin{eqnarray}\nonumber
\widehat{H}&=&\hbar\omega(\hat{a}^{\dagger}_1\hat{a}_1+\hat{a}^{\dagger}_2\hat{a}_2)+\hbar\Omega\hat{b}^{\dagger}\hat{b}\\ \label{OMSHamiltonian}
&&-\hbar g_0(\hat{a}^{\dagger}_1\hat{a}_1-\hat{a}^{\dagger}_2\hat{a}_2)(\hat{b}^{\dagger}+\hat{b}).
\end{eqnarray}
The operators $\hat{a}_i$, $i=1,2$, are the (boson) cavity mode operators of each side of the cavity (the side $i=1$ being the upper side and $i=2$ the lower side). Both optical modes have the same frequency $\omega$. The operator $\hat{b}$ is the annihilation operator of excitations of the center of mass of the mirror, which is treated as a harmonic oscillator of frequency $\Omega$. The first two terms (from the left to the right) represent the free energy of the cavity and the mechanical oscillator, while the last term constitutes the optomechanical interaction via radiation pressure. The parameter $g_0$ is the \emph{vacuum optomechanical coupling strength} that characterises the magnitude of the coupling between a single photon and a single phonon. In our setup 
\begin{eqnarray}\label{g0}
g_0=\Big(\frac{\omega}{L}\Big)x_0,
\end{eqnarray}
where $L$ is the effective length of each side of the cavity and $x_0$ are the  \emph{zero-point fluctuations} of the mechanical oscillator, which in turn are defined as  
\begin{eqnarray}\label{x0}
x_0=\sqrt{\frac{\hbar}{2M\Omega}},
\end{eqnarray}
where $M$ is the mass of the mirror. For a mirror of the size of the order of $\mu$m, $x_0\sim 10^{-15}$ m. Thus, for an optical cavity with length $L\sim$ mm, $g_0\sim1-10^{3}$ Hz. 

The evolution operator $\widehat{U}_H(t)=\exp{\{-(i/\hbar)\widehat{H}t\}}$ generated by the hamiltonian (\ref{OMSHamiltonian}) can be factorized into different operators. Following a calculation analogous to the one presented in \cite{Bose1997} it is easy to show that $\widehat{U}_H(t)=\widehat{U}_{cav}(t)\times\widehat{U}_{Kerr}(t)\times\widehat{U}(t)\times\widehat{U}_M(t)$, where 
\begin{eqnarray}\nonumber
\widehat{U}_{cav}(t)&=&\exp{\{-i\omega t(\hat{a}^{\dagger}_1\hat{a}_1+\hat{a}^{\dagger}_2\hat{a}_2)\}},\\ \nonumber
\widehat{U}_{Kerr}(t)&=&\exp{\{ig^2\cdot\phi(t)(\hat{a}_1^{\dagger}\hat{a}_1-\hat{a}_2^{\dagger}\hat{a}_2)^2\}},\\ \nonumber
\widehat{U}(t)&=&\exp{\{g\cdot(\hat{a}^{\dagger}_1\hat{a}_1-\hat{a}^{\dagger}_2\hat{a}_2)\cdot[\hat{b}^{\dagger}\varphi(t)-\hat{b}\varphi^*(t)]\}},\\
\widehat{U}_M(t)&=&\exp{\{-i\Omega t\hat{b}^{\dagger}\hat{b}\}}.
\end{eqnarray}
The operators $\widehat{U}_{cav}(t)$ and $\widehat{U}_{M}(t)$ represent the free evolution of the cavity and mechanical modes, respectively.  The term $\widehat{U}_{Kerr}(t)$ adds a phase that depends quadratically on the difference of photons between both sides of the cavity, which shows that the optomechanical coupling generates an effective Kerr nonlinearity or photon-photon interaction. The term $\widehat{U}(t)$ generates optomechanical entanglement between  ``microscopic'' degrees of freedom (the cavity modes) and ``macroscopic'' degrees of freedom (the center of mass of a vibrating mirror). The time-dependent functions $\phi(t)$ and $\varphi(t)$ ($\varphi^*(t)$ denotes its complex conjugate) are defined as
\begin{eqnarray}\nonumber
\phi(t)&=&\Omega t-\sin(\Omega t),\\ 
\varphi(t)&=&1-e^{-i\Omega t}.
\end{eqnarray}
The adimensional parameter $g$ that appears in the expressions for $\widehat{U}_{Kerr}(t)$ and $\widehat{U}(t)$ is defined as $g\equiv g_0/\Omega$. A physical interpretation of $g$ will be given later. 

The hamiltonian (\ref{OMSHamiltonian}) preserves the total number of photons $\widehat{N}=\hat{a}_1^{\dagger}\hat{a}_1+\hat{a}_2^{\dagger}\hat{a}_2$, i.e. $[\widehat{N},\widehat{H}]=0$. Therefore, a state with a well defined number of photons will remain in this subspace under the action of $\widehat{U}_H(t)$. In our experiment we will use single photons and, consequently, the optical component of the Hilbert space of states will be reduced to the single photon subspace. Restricted to this subspace it is possible to define the operators 
\begin{eqnarray}\nonumber
\hat{\sigma}_x&=&\hat{a}^{\dagger}_1\hat{a}_2+\hat{a}^{\dagger}_2\hat{a}_1,\\ \nonumber
\hat{\sigma}_y&=&(\hat{a}^{\dagger}_1\hat{a}_2-\hat{a}^{\dagger}_2\hat{a}_1)/i, \\ \nonumber
\hat{\sigma}_z&=&\hat{a}^{\dagger}_1\hat{a}_1-\hat{a}^{\dagger}_2\hat{a}_2,\\ \label{SchwingerMap}
\widehat{N}&=&\mathbb{1},
\end{eqnarray}
which satisfy $[\hat{\sigma}_x,\hat{\sigma}_y]=i\hat{\sigma}_z$ (and its cyclic permutations), and $\hat{\sigma}_x^2=\hat{\sigma}_y^2=\hat{\sigma}_z^2=\mathbb{1}$. This map between operators is called the Jordan-Schwinger map \cite{Schwinger,Yurke}. Hence, the operators $\widehat{U}_{cav}(t)$ and $\widehat{U}_{Kerr}(t)$ will merely add a global phase factor and may be disregarded. Moreover, as will be explained in the next section, the mechanical oscillator will be initialized in the ground state and thus the free evolution of the mirror $\widehat{U}_M(t)$ will play no role. Thereby, under these conditions (when the mirror starts in the ground state and single photons are employed) the evolution operator $\widehat{U}_H(t)$ will be simply reduced to 
\begin{eqnarray}\label{SimplifiedEvolutionOperator}
\widehat{U}(t)=\exp{ \{g \cdot \hat{\sigma}_z \cdot [\hat{b}^{\dagger}\varphi(t)-\hat{b}\varphi^*(t)]\}}.
\end{eqnarray}

The time $t$ spent by one photon inside the cavity before it is emitted into the environment is not a deterministic time, but a random time that follows an exponential distribution, i.e. $t\sim\exp{-\gamma t}$, where $\gamma$ is rate at which the energy is dissipated from each side of the cavity (assumed to be equal in both sides). Therefore, the average time spent by one photon will be $\gamma^{-1}$. It will be assumed that, on average, the photon stays for half a mechanical period inside the OMS before it is emitted, i.e. $\Omega t=\pi$ (which can be achieved by making $\gamma=\Omega/\pi$).  The reason for this consideration is that at half a vibrational period the displacement of the position of the oscillator produced by a single photon reaches a maximum value. Consequently, the evolution operator (\ref{SimplifiedEvolutionOperator}) becomes
\begin{eqnarray}\label{SimplifiedEvolutionOperator2}
\widehat{U}=\exp{ \{-(i/\hbar)\cdot  (4gx_0) \cdot \hat{\sigma}_z \widehat{P}\}}.
\end{eqnarray}
The operator $\hat{P}$ is the momentum of the center of mass of the oscillator. Notice that the evolution operator has the form a measurement, according to the so called \emph{von Neumann model} \cite{Svensson2013,Kofman2012,VonNeumann}. Indeed, a discrete system variable $\hat{\sigma}_z$ couples to a continuous variable $\widehat{P}$ of the measurement device (the mirror). The result of the measurement of $\hat{\sigma}_z$ is obtained by observing the conjugate variable to $\hat{P}$, i.e. by observing the position $\widehat{Q}$ of the mirror. 

From (\ref{SimplifiedEvolutionOperator2}) it is clear that the parameter $g$ corresponds to the displacement produced by a single photon in the (mechanical) phase space, i.e. it is the displacement of the equilibrium position of the oscillator in units of $x_0$.  For an oscillator with frequency $\Omega\sim 1-100$ MHz, $g$ is a small parameter with values in a wide range, $10^{-8}-10^{-2}$. We will be interested in the estimation of this parameter. 

\subsection{OMS in Mach-Zehnder Interferometer}\label{sec:sec2:sub2} 

It will be assumed that the mirror starts cooled down to its ground state, i.e.  $\hat{\rho}_M=\ket{0}\bra{0}$ (the subscript $M$ indicates that the state corresponds to a state of the measurement device). The position representation of $\ket{0}$ is given by
\begin{eqnarray}
\ket{0}=\int dq \Big(\frac{1}{\sqrt{2\pi}x_0}\Big)^{1/2}  \exp{ \Big(  - \frac{q^2}{4x_0^2} \Big) } \ket{q}.
\end{eqnarray}
Also, the electromagnetic field inside the cavity should begin in a pure  single photon state, namely,  $\rho_S=\ket{i}\bra{i}$ (the subscript $S$ indicates that the state corresponds to a state of the system). The details regarding the preparation of $\hat{\rho}_S$ will be given later. 

Therefore, the initial state of the system (S) and the measurement device (M) is given by
\begin{eqnarray}
\hat{\rho}_{SM}(0)=\hat{\rho}_S\otimes\hat{\rho}_M.
\end{eqnarray}
The state $\hat{\rho}_{SM}(0)$ will evolve under the action of (\ref{SimplifiedEvolutionOperator2}). Hence, the state of the OMS immediately after the photon has been emitted will be 
\begin{eqnarray}
\hat{\rho}_{SM}=\widehat{U}\hat{\rho}_{SM}(0)\widehat{U}^{\dagger}.
\end{eqnarray} 

Now, we will distinguish two measurement strategies of $\hat{\sigma}_z$: i) \emph{without postselection} of photons (NPS), and ii) \emph{with postselection} (PS). In the first strategy we will be interested in the average position of the measurement device, \emph{independently} of the photon counter at which the photon is detected after it is released from the OMS.  In the second strategy, we will focus on the average position of the measurement device \emph{conditioned} on the detection of the photon in the so called dark port (see Fig. \ref{fig:fig1}). 

The experiment will be repeated in $N$ independent trials. Since in each trial one single photon is employed, in total, $N$ quantum resources will be used.  We analyze now both strategies. 

\subsubsection{NPS measurement strategy}\label{NPS strategy} The average position of the measurement device using the NPS strategy is given by
\begin{eqnarray}
\langle{\hat{Q}}\rangle_{NPS}=\Tr (\hat{Q}\hat{\rho}_{SM})=4gx_0\langle \hat{\sigma}_z\rangle,
\end{eqnarray}
where $\langle \hat{\sigma}_z\rangle=\bra{i} \hat{\sigma}_z\ket{i}$. Thus, in order for the displacement to be maximum, the initial state of the cavity should be
\begin{eqnarray}
\ket{i}=\ket{1,0}\quad\text{or}\quad\ket{i}=\ket{0,1}.
\end{eqnarray}
The first entry of $\ket{\text{ },\text{ }}$ represents the number of photons in the mode of the side $1$ of the the cavity, while the second entry corresponds to photons in the other mode. These states can be prepared by removing BS1 and injecting the single photon directly through one of the arms of the interferometer. Then, if the photon is properly prepared, and in the absent of any looses, it will be absorbed by the cavity.  Thus, the displacement with the NPS strategy will be $\langle{\hat{Q}}\rangle_{NPS}=4gx_0$. 

\subsubsection{PS measurement strategy}\label{PS strategy}

When using postselection a single photon is sent into the interferometer through one of the input ports, while the other port is leaved unused (in the vacuum state). The photon will enter the interferometer through BS1, which separates the light into a reflected component (propagating through the arm $1$) and a transmitted component (propagating through the arm $2$). Both components have equal intensities since the beam splitter is balanced. The single photon state is a highly non classical state and thus, inside I1, a path entangled state will be generated, i.e. a coherent superposition of the photon propagating along each arm. Therefore, after the photon has been absorbed by the OMS and assuming no losses, the initial state of the light inside the OMS will be 
\begin{eqnarray}
\ket{i}=\frac{1}{\sqrt{2}}\ket{1,0}+\frac{1}{\sqrt{2}}\ket{0,1}.
\end{eqnarray}

On the other hand,  the pure state $\ket{f}$ is the postselected state when the dark port ``clicks'', and is given by 
\begin{eqnarray}
\ket{f}=t\ket{1,0}-r\ket{0,1}.
\end{eqnarray}
The coefficients $t$ and $r$ are the (real and positive) transmittances an reflectances of BS2, respectively. In ideal conditions the beamsplitter preserves the energy, i.e. $t^2+r^2=1$. Hence, $0\leq r \leq 1$ and $0\leq t \leq 1$. See Appendix \ref{app:A} for details regarding the quantum mechanical description of  BS2.  It will be useful to define a parameter $\delta$ that quantifies the level of unbalance of BS2 as
\begin{eqnarray}\label{ParameterUnbalance}
\delta=t-r.
\end{eqnarray}
When $\delta=0$ BS2 is a balanced beamsplitter. The configuration $\delta=1$ corresponds to the scenario in which BS2 is removed, while when $\delta=-1$ BS2 behaves as a perfectly reflecting mirror. 

Under these considerations the conditioned state of the measurement device is given by 
\begin{eqnarray}\label{MPS}
\hat{\rho}_{M,PS}=\frac{\bra{f}\hat{\rho}_{SM}\ket{f}}{\Tr(\bra{f}\hat{\rho}_{SM}\ket{f} )}.
\end{eqnarray} 
The term in the denominator of (\ref{MPS}) corresponds to the probability $p$ to successfully detect a photon in the dark port, and corresponds to
\begin{eqnarray}
p=\frac{1-e^{-8g^2}}{2}+\frac{\delta^2}{2}e^{-8g^2}.
\end{eqnarray}
Since $g\ll1$ we can safely expand the exponential term to second order, 
\begin{eqnarray}
p=4g^2+\frac{\delta^2}{2}.
\end{eqnarray}
This expression allows to define two regimes that will be important for the our analysis: i) the weak measurement regime ($\delta\gg g$) and, ii) the strong measurement regime ($\delta \sim g$). 

The average position of measurement device using the PS strategy corresponds to
\begin{eqnarray}
\langle{\hat{Q}}\rangle_{PS}=\Tr (\hat{Q}\hat{\rho}_{M,PS})=4gx_0 \Bigg( \frac{\delta\sqrt{2-\delta^2}}{2p}\Bigg).
\end{eqnarray}
In the weak measurement regime, this expression reduces to $\langle{\hat{Q}}\rangle_{PS}=4gx_0 \sigma_{z,w}$, where $\sigma_{z,w}$ is the \emph{quantum weak value} of the operator $\hat{\sigma}_z$ between the initial state $\ket{i}$ and the final state $\ket{f}$,
\begin{eqnarray}
\sigma_{z,w}=\frac{\sqrt{2-\delta^2}}{\delta}.
\end{eqnarray}
For $|\delta|\neq1$ the weak value is \emph{anomalous} in the sense that it is larger, in magnitude, than any of the eigenvalues of $\hat{\sigma}_z$, and gets larger as $|\delta|\rightarrow0$. Recall, however, that the weak measurement regime is restricted by $\delta\gg g$. 

\section{Fisher Information Analysis}\label{sec:sec3}

The average position of the measurement device will be recorded using I2, in which a classical beam with $|\alpha|^2$ photons is sent through one of the input ports while the other is in the vacuum state. The laser has the same frequency as the frequency of the cavity ($\omega$). The operation of I2 will be triggered $M$ times. When the NPS strategy is employed $M=N$, i.e. I2 will be operated for every single photon, independently on which of the two detectors of I1 clicks.  For the PS strategy, I2 will be activated only when a single photon is detected in the dark port of I1. Therefore, in this case, $M=Np$. 

In each operation of I2 the difference of photons recorded at the detectors located in the output ports of I2 (normalized by $|\alpha|^2$) will be read. Therefore, in total, there will be $M$ readings, 
\begin{align}\label{ResultsI2}
    \vec{R} = \begin{pmatrix}
          R_1\\
          \vdots\\
        R_M\\
         \end{pmatrix}.
  \end{align}
The random vector $\vec{R}$ will follow a multivariate normal distribution, 
\begin{eqnarray}\label{MultivariateGaussianDistrib}
f_{\vec{R}}(\vec{r}|g)=\frac{1}{\sqrt{2\pi|C|}}\exp{-\frac{(\vec{r}-\vec{\mu})^TC^{-1}(\vec{r}-\vec{\mu})}{2}},
\end{eqnarray}
where $\vec{r}$ is a $M$-dimensional vector whose elements are $r_i$. The vector $\vec{\mu}$ is the vector of $M$ means, all of which are identical and proportional to the length difference between both paths of I2, i.e. $\vec{u}=2(\omega/c)\cdot\langle \hat{Q}\rangle_k \cdot \mathbb{1}$, where $k=NPS, PS$, $\mathbb{1}$ is the $M$-dimensional identity vector, and $c$ is the speed of light.    

Also, $C$ is the $M\times M$ covariance matrix, $C^{-1}$ is its inverse matrix, and $|C|$ stands for the determinant. The form of the covariance matrix defines the type of noise that affects the measurement. Here, we will consider the model of correlated noise 
used in \cite{Feizpour2011,Sinclair2017}, which is characterized by a covariance matrix with elements $C_{i,j}=|\alpha|^{-2}[\delta_{i,j}+\gamma(1-\delta_{i,j})]$, namely, 
\begin{align}\label{CovarianceMatrix}
    C=|\alpha|^{-2} \begin{pmatrix}
          1&\gamma&\hdots&\gamma  \\
          \gamma&1& &\\
          \vdots& &\ddots& \\
           \gamma& & & 1 \\
         \end{pmatrix}.
  \end{align}
The parameter $0\leq\gamma\leq1$ measures the amount of correlation between the different readings. If $\gamma=0$ the noise is said to be uncorrelated or \emph{white noise}, and is said to be correlated or \emph{colored noise} in any other case.  

The classical \emph{Fisher information} \cite{Fisher} with respect to the parameter $g$ contained in the probability distribution (\ref{MultivariateGaussianDistrib}) is defined as 
\begin{eqnarray}\label{FisherInformation}
\mathcal{I}(g)=\Big\langle -\frac{\partial^2}{\partial g^2}\ln f_{\vec{R}}(\vec{r}|g) \Big\rangle,
\end{eqnarray}
where the average is taken over the distribution (\ref{MultivariateGaussianDistrib}). The classical Fisher information defines a lower bound on the variance of any unbiased estimator $\hat{g}$ of the parameter $g$. This lower bound is called the \emph{Cram\'{e}r-Rao bound} \cite{Cramer,Kay2010,VanTrees}, 
\begin{eqnarray}
\langle\Delta \hat{g}^2\rangle \geq \mathcal{I}^{-1}(g).
\end{eqnarray}
The variance $\langle\Delta \hat{g}^2\rangle$ is a measure of the precision of the estimator (among two unbiased estimators the one with lower variance is preferable). Therefore, and according to the Cram\'{e}r-Rao bound, the larger the Fisher information, the smaller will be the lowest attainable variance of any unbiased estimator of $g$.

Using the definition (\ref{FisherInformation}) it is possible to obtain exact expressions for the Fisher information for each measurement strategy, because the noise model (\ref{CovarianceMatrix}) is exactly solvable. The Fisher information for each measurement strategy corresponds to 
\begin{eqnarray}\label{FisherPS}
\mathcal{I}_{PS}(g)&=&I_0\Big(\frac{N}{1+Np\gamma}\Big)h(\delta,g),\\ \label{FisherNPS}
\mathcal{I}_{NPS}(g)&=&I_0\Big(\frac{N}{1+N\gamma}\Big).
\end{eqnarray}
The value  $I_0=|\alpha|^2\cdot(8x_0\omega/c)^2$ depends on the number of photons contained in the classical beam and to avoid additional factors we will assume that the laser has enough power so that $I_0\sim1$. The factor $h(\delta,g)=\delta^2\cdot(2-\delta^2)\cdot(\delta^2-24g^2)^2/(16p^3)$ depends both on $g$ (the unknown parameter we wish to estimate) and $\delta$, and it is bounded as $0\leq h(\delta,g)\leq1$. 

\subsection{White Noise}

In the white noise case  $\gamma=0$. Thus, the Fisher information (\ref{FisherPS})  and (\ref{FisherNPS}) will scale as $N$ (SQL). The only difference between both strategies is the factor $h(\delta,g)$, which needs to be close to unity (and not near zero) in order to minimize the difference. 

In the weak measurement regime ($\delta \gg g$), this factor reduces to
\begin{eqnarray}\label{FactorWMR}
h(\delta,g)=1-\frac{\delta^2}{2}=p\cdot\sigma_{z,w}^2. 
\end{eqnarray}
The square of the weak value $\sim p^{-1}$ which shows that amplification effect is exactly canceled by the lesser amount of operations of I2 (as compared to the NPS strategy). Expression 
(\ref{FactorWMR}) also shows that in the regime $1\gg\delta\gg g$  the PS strategy (which relies on weak values) produces the same Fisher information as the NPS strategy, i.e. both strategies have the same maximum precision for the estimation of $g$. Using the first strategy, however, the amount of operations of I2 is reduced from $N$ to $Np=N\delta^2/2$. 

\begin{figure}
 \centering \includegraphics[width=\linewidth]{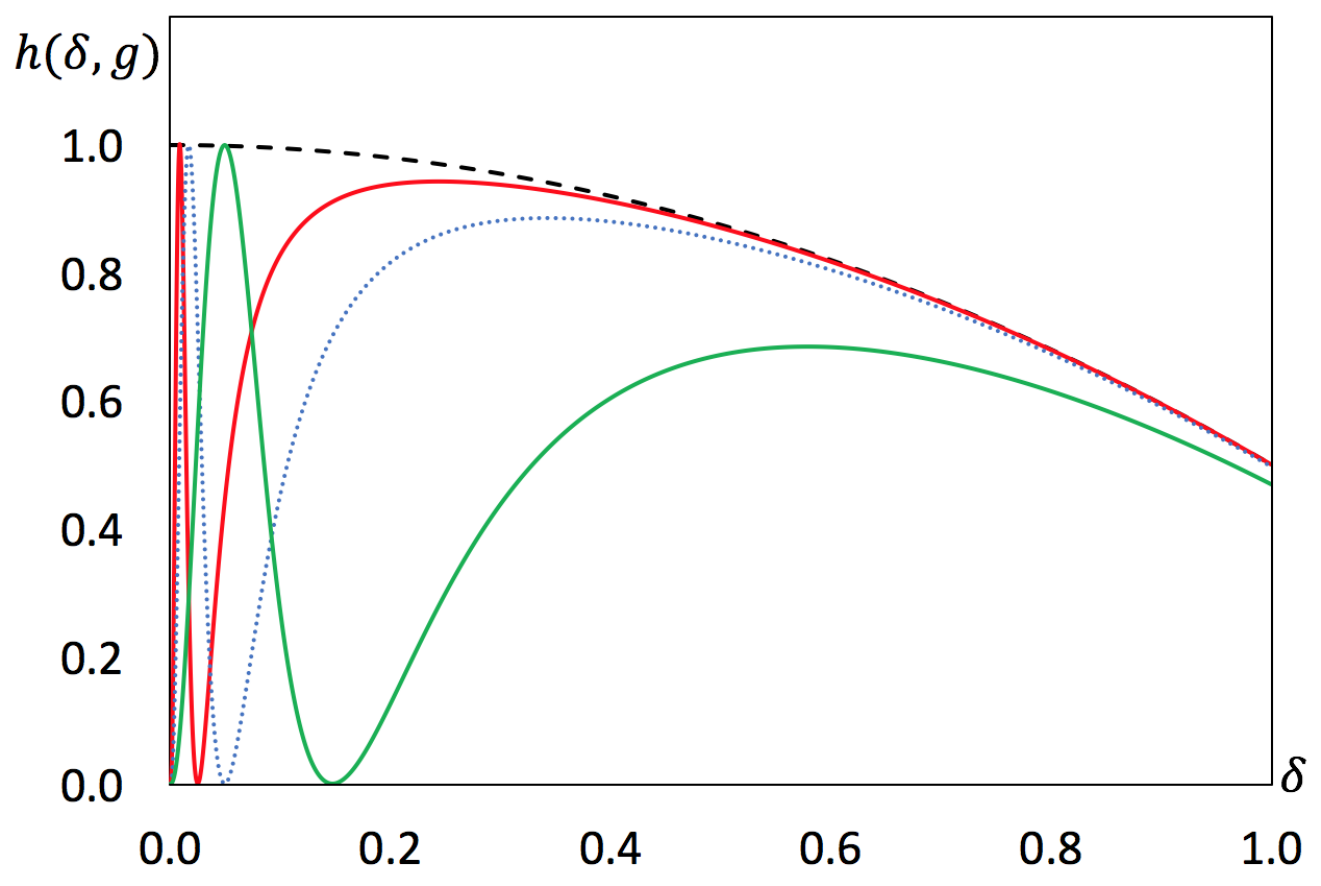}
 \caption{The factor $h(\delta,g)$ is plotted against the postselection parameter $\delta$, considering $g=5\cdot10^{-3}$ (red curve), $g=10^{-2}$ (blue dotted curve) and $g=2\cdot10^{-2}$ (green curve). In the weak measurement regime these curves coincide with the black dashed curve (which corresponds to $1-\delta^2/2$). Notice that when $\delta\sim g$ each curve reaches a maximum value close to unity but quickly drops to zero in the vicinity at which the maximum is reached. This shows that for small values of $g$ the weak measurement regime is more convenient since it admits a larger interval of possible values of $\delta$, i.e. it does not require precise knowledge of the parameter.} \label{fig:fig4}
\end{figure}

On the other hand, when  the measurement is strong ($\delta \sim g$), the factor is close to unity, but it quickly drops to zero, as can be seen in Fig. \ref{fig:fig4}. Therefore, a useful measurement strategy that relies on  postselection and a strong measurement would require good prior knowledge of the very parameter we wish to estimate. When using weak measurements, on the contrary, we only have to know that $\delta\gg g$. 

\subsection{Colored Noise}

\begin{figure}
 \centering \includegraphics[width=\linewidth]{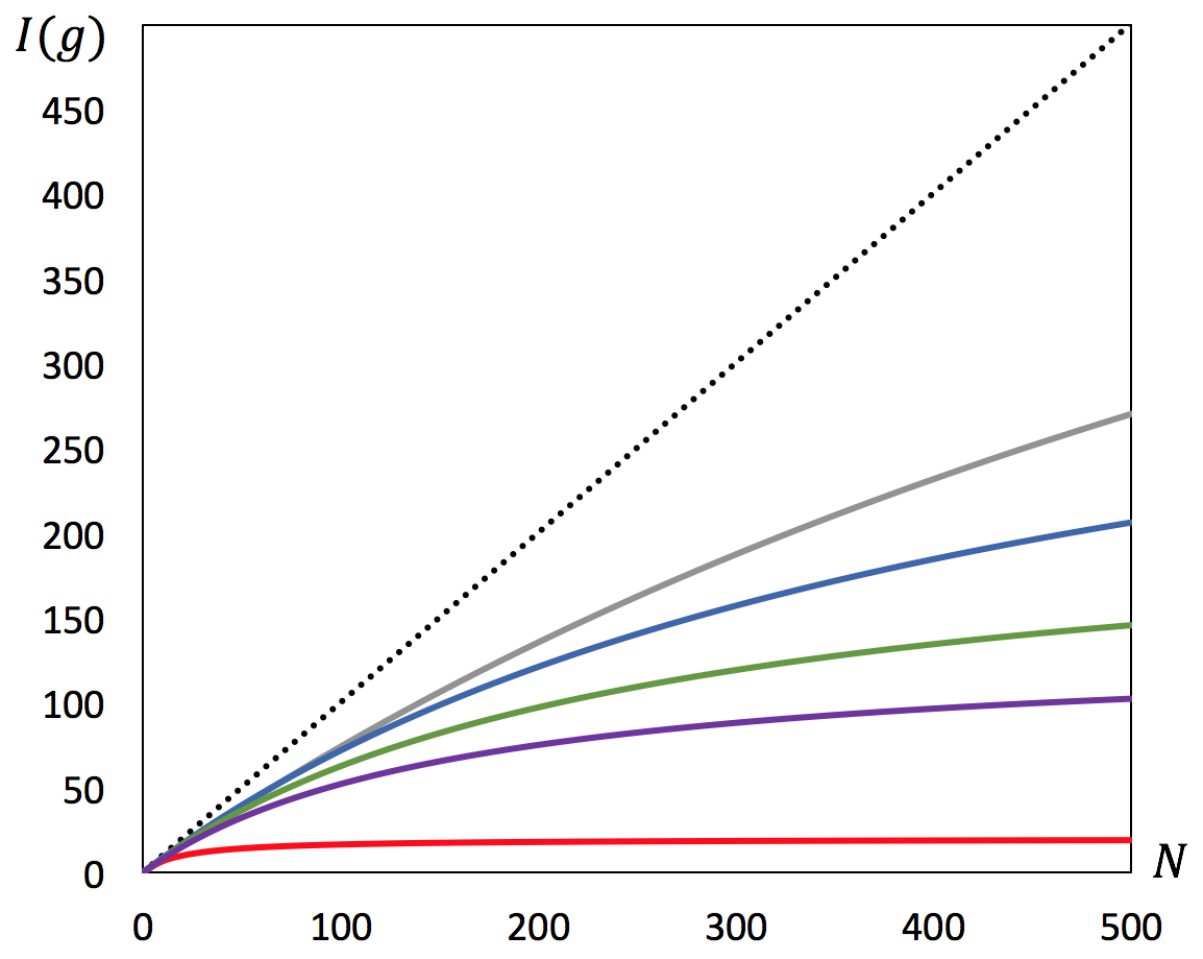}
 \caption{The Fisher information $\mathcal{I}(g)$ is plotted as a function of the number of single photons $N$ for different values of the postselection parameter $\delta$; grey curve ($\delta=0.2, p=2.0\%$ and $\sigma_{z,w}=7.0$), blue curve ($\delta=0.3, p=4.5\%$ and $\sigma_{z,w}=4.6$), green curve ($\delta=0.4, p=8.0\%$ and $\sigma_{z,w}=3.4$), and magenta  curve ($\delta=0.5, p=12.5\%$ and $\sigma_{z,w}=2.6$). The red curve corresponds to the scenario without postselection. Initially, in all cases, the Fisher information scales linearly and for large $N$ reaches a constant value equal to $(p\gamma)^{-1}$ when postselection is employed and equal to $\gamma^{-1}$ when no postselection is taken into account. For all curves $g=10^{-2}$ and  $\gamma=5\cdot 10^{-2}$. } \label{fig:fig5}
\end{figure}

When the noise is correlated the parameter $\gamma$ will be different from zero. In this case, by comparing (\ref{FisherPS}) and (\ref{FisherNPS}), and as it is pointed out in \cite{Sinclair2017}, it is clear that the size of the correlations $\gamma$ is reduced by a factor $p$ when postselection is employed. Consequently, the factor in brackets that appears in (\ref{FisherPS}) will be larger than the one appearing in (\ref{FisherNPS}) and, thereby, the Fisher information with the PS strategy will be larger, as long as the factor $h(\delta,g)$ does not vanish. As in the previous case, this fact can be achieved by using a weak measurement with postselection, and having the certainty that $g\ll\delta$ (a strong measurement would require precise knowledge about $g$). Fig. \ref{fig:fig5} shows the benefits of using postselection when correlated noise affects the measurement, for different values of the postselection parameter $\delta$. 

Finally, it is worth to mention that in this scenario the Fisher information initially scales as $N$ but then saturates at a value equal to the inverse of the size of the correlations, i.e. at $\gamma^{-1}$ and $(\gamma p)^{-1}$ without and with postselection, respectively. 

\section{Discussion}\label{sec:sec4} 

In this article this we have applied the results regarding weak measurements with postselection in the presence of correlated noise developed in \cite{Feizpour2011,Jordan2014,Sinclair2017} to the estimation of optomechanical parameters.  We have presented an experimental proposal in which dark port postselection of single photons, together with anomalous weak values, allows to reach the same precision as with a strategy without postselection, in the presence of white noise. In this case, the larger the weak value (i.e. the smaller the probability of postselection), the smaller the difference between both protocols. When the noise has correlations (colored noise) the maximum precision is increased with postselection. As in the previous case, the larger the weak value, the better the maximum precision. 

In both scenarios, either with uncorrelated or correlated noise, the weak value is restricted by $\delta\gg g$ and thus can not be made indefinitely large. In the strong measurement regime a good prior knowledge of the parameter we aim to estimate is required (otherwise the Fisher information may be close to zero) and, consequently, weak measurements are preferable over strong measurements. 

The parameter $g$ may be estimated with the maximum likelihood estimator (MLE) \cite{VanTrees} using the distribution (\ref{MultivariateGaussianDistrib}). For the PS strategy based on weak values the MLE corresponds to 
\begin{eqnarray}
\hat{g}=\frac{1}{8x_0(\omega/c) \sigma_{z,w}} \frac{\sum_{i=1}^{Np} R_i}{Np}. 
\end{eqnarray}
The variance of the MLE is given by
\begin{eqnarray}\label{PSMLE}
\langle \Delta \hat{g} ^2 \rangle=\Big[\frac{1+\gamma(Np-1)}{I_0 N}\Big]\frac{1}{\sigma_{z,w}^2p},
\end{eqnarray}
which shows that in general the MLE reaches asymptotically (for large $N$) the maximum precision defined by the Cram\'{e}r-Rao bound, $\mathcal{I}_{PS}^{-1}(g)$, i.e. it is asymptotically efficient (which is a property of this kind of estimator).  The estimator is independent of $g$ in the regime $g\ll\delta\ll1$, and note also that when the noise is white the MLE is efficient for every value of $N$ (it saturates exactly the Cram\'{e}r-Rao bound). The MLE for the strategy without postselection is analogous, 
\begin{eqnarray}\label{NPSMLE}
\hat{g}=\frac{1}{8x_0(\omega/c) } \frac{\sum_{i=1}^{N} R_i}{N}, 
\end{eqnarray}
and is asymptotically efficient for correlated noise and efficient for every sample size when the noise is white. 

As can be seen from (\ref{NPSMLE}) when using the NPS strategy all data is treated equally, i.e. there is not distinction between the data coming from the ensemble of successfully postselected photons and the data coming from the group of photons detected in the bright port. On the contrary, when the PS strategy is employed, the non postselected photons are simply disregarded (the operation of I2 is not triggered in those cases) while all the data comes only from the ensemble of postselected photons. In fact, the estimator (\ref{NPSMLE}) is constructed only from observations triggered when photons are successfully detected in the dark port. Certainly, there are other strategies that take into account both ensembles and threat them differently, for which the Fisher information will be (possibly slightly) increased \cite{Ferrie2013,Sinclair2017}. The PS strategy however exhibits the benefit of triggering I2 fewer times and reaching the same level of maximum precision for the estimation.  

\section*{ACKNOWLEDGMENTS}
We thank the financial support of ANID with the project Fondecyt $\#$1180175.

\appendix

\section{Quantum Mechanical Description of BS2}\label{app:A}

\begin{figure}
 \centering \includegraphics[width=0.35\textwidth]{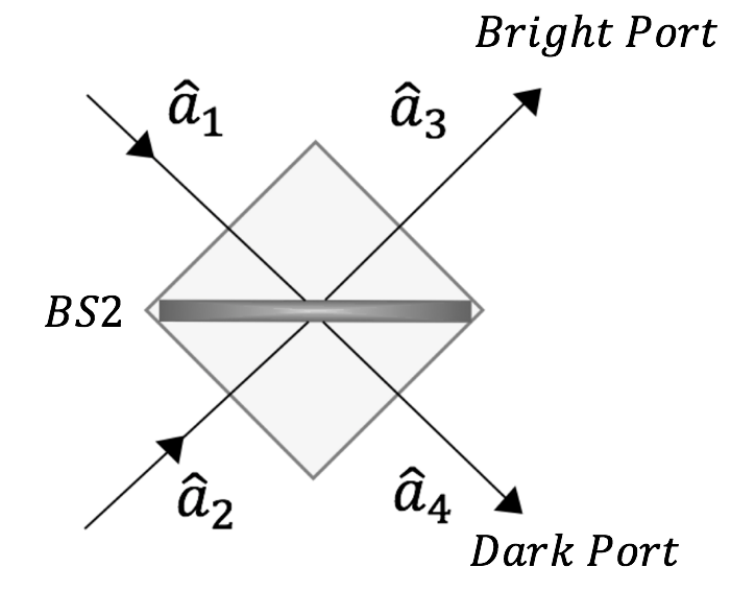}
 \caption{BS2: unbalanced beam splitter located at the output of I1 where the modes $\hat{a}_1$ and $\hat{a}_2$ are coherently mixed. The parameter $\delta=t-r$ quantifies the level of unbalance between the transmittance $t$ and the reflectance $r$. 
 } \label{fig:fig6}
\end{figure}

Consider the description of BS2 shown in Fig. \ref{fig:fig6}. The relationship between the input and output modes of BS2 is given by 
\begin{eqnarray}\nonumber
\hat{a}_3&=&r\hat{a}_1+t\hat{a}_2,\\ \label{UnbalancedBS}
\hat{a}_4&=&t\hat{a}_1-r\hat{a}_2,
\end{eqnarray}
where $t$ and $r$ are (real and positive) transmittances and reflectances of the beamsplitter, respectively. The minus sign in the second equation is necessary in order for the transformation to be unitary. Physically, it means that the field reflected in lower side of the beamsplitter acquires a phase of $\pi$. The transformation of the fields (\ref{UnbalancedBS}) allows to transform states between the inner and outer paths of the interferometer, as it is described in the following example.
\begin{eqnarray}\nonumber
\ket{0}_3\ket{1}_4&=&\hat{a}_4^{\dagger}\ket{0}_3\ket{0}_4\\ \nonumber
&=&(t\hat{a}_1-r\hat{a}_2)\ket{0}_1\ket{0}_2\\  \label{TransformationUBS}
&=&t\ket{1}_1\ket{0}_2-r\ket{0}_1\ket{1}_2.
\end{eqnarray}
The subindices of the states denote the corresponding paths in the interferometer. Notice that in the second step of (\ref{TransformationUBS}) the beam splitter transformation (\ref{UnbalancedBS}) was employed, while the vacuum states outside and inside the interferometer were identified. Consequently, counting a photon in the arm $4$ is equivalent to select the state 
\begin{eqnarray}\label{PSstateDark}
\ket{\psi_f}=t\ket{1}_1\ket{0}_2-r\ket{0}_1\ket{1}_2
\end{eqnarray}
inside the interferometer.

\end{document}